\documentclass[pdflatex,sn-mathphys-num,margin=1in]{sn-jnl}
\usepackage{graphicx,multirow, mathrsfs, amsthm, amsmath,amssymb,amsfonts}%
\usepackage[title]{appendix}%
\usepackage{textcomp}%
\usepackage{manyfoot}%
\usepackage{booktabs}%
\usepackage{algorithm}%
\usepackage{algorithmicx}%
\usepackage{algpseudocode}%
\usepackage{listings}%
\usepackage[table,xcdraw]{xcolor}
\usepackage{pdflscape} 
\theoremstyle{thmstyleone}%
\theoremstyle{thmstyletwo}%
\theoremstyle{thmstylethree}%
\begin{document}
\title{Multi-polarization during the 9th European Parliament}
\author[1]{\fnm{Sebastião M.} \sur{Rosalino}} \email{sorosalino@gmail.com}
\author[1]{\fnm{António} \sur{Curado}}\email{antonio.b.curado@gmail.com}
\author[1]{\fnm{Flávio L.} \sur{Pinheiro}}\email{fpinheiro@novaims.unl.pt}

\affil[1]{\orgdiv{NOVA Information Management School (NOVA IMS)}, \orgname{Universidade Nova de Lisboa}, \orgaddress{\street{Campus de Campolide}, \city{Lisboa}, \postcode{1070-312}, \country{Portugal}}}


\abstract{Brexit, a global pandemic, the Russian invasion of Ukraine, and record inflation – few legislative bodies have faced such a cascade of shocks as the European Parliament did during its 9th term (2019-2024). Using the Bipartite Configuration Model and a set of network statistics, this research explores how multi-polarization was characterized during this term by constructing and analyzing co-voting networks across all legislative subjects and within specific legislative subjects. The results contest binary polarization narratives inherited from US/UK scholarship by uncovering a multi-polar landscape. In many legislative subjects, including “Community policies”, “Internal market, single market”, and “External relations of the Union”, coalitions realign fluidly, forming several voting communities rather than a single left-right divide. Ideological affinity and group memberships, not nationality, emerge as the primary forces that bind or separate Members of the European Parliament, reaffirming the chamber’s transnational character. Two quantitative patterns stand out. First, the Greens/EFA and The Left display the highest intragroup cohesion, while governing groups -- EPP, S\&D, and Renew -- often fracture into multiple, issue-driven alliances, suggesting declining centrist disciplines. Second, a distinct Eurosceptic versus Euroenthusiastic cleavage crystallizes in matters concerning the “State and evolution of the Union” subject, cutting across economic and social ideologies and hinting at a budding second dimension of parliamentary conflict. This research highlights how legislative consensus in the European Parliament will hinge on navigating a fluid, multi-polar, issue-driven alliance landscape rather than building stable grand coalitions.}


\maketitle

\section{Introduction}
Political and social polarization are among the most pressing challenges facing liberal democracies today \cite{dixit2007political,conover2011political,schedler2023rethinking}. Polarized societies become entrenched in ideological silos and governance becomes a battleground where collaboration is sacrificed for partisanship. Indeed, while polarization can be a catalyst for political engagement and spur public discourse \cite{Rogowski2016}, it frequently creates irreconcilable divisions that impede decision-making and erode trust in democratic institutions \cite{McCarty2016}. This phenomenon has become a defining feature of the global political landscape, influencing electoral dynamics, legislative processes, and public policy outcomes \cite{Iyengar2015}.

In Germany, France, and Italy, political polarization led to the emergence of new parties and the fragmentation of traditional parliamentary structures and coalitions. In the United States, political polarization is characterized by a stark ideological bifurcation within established parties, with Democrats tending toward greater progressivism and Republicans embracing heightened conservatism \cite{Canen2020, Patks2023}.

The intensification of political polarization in recent decades must be understood not merely as a shift in partisan preferences but as a structural transformation driven by a complex interaction of economic, technological, and demographic forces. Polarization is often linked to widening income inequality, technological disruptions in labor markets, and the emergence of fragmented media ecosystems that reinforce ideological silos \cite{Autor2020, Grechyna2016, Iyengar2012}. These drivers have contributed not only to ideological polarization—manifested in the distance between party platforms—but also to affective polarization, wherein opposing political groups are perceived not just as adversaries, but as threats to social identity and cohesion \cite{Iyengar2015, Druckman2019}.

Financial and migratory crises have further exacerbated this dynamic, particularly in Europe, where increased immigration flows and austerity-driven responses to sovereign debt have fueled support for anti-establishment parties. The 2008 global financial crisis, for instance, acted as a catalyst for the rise of radical-right populist movements, undermining centrist coalitions and intensifying identity-based cleavages \cite{Funke2016, Rodrik2018}. Similarly, evidence suggests that in regions facing high immigration inflows, electoral support for far-right parties has increased in direct response to perceived cultural and economic threats \cite{Otto2014, Halla2017}.

Within this context, affective polarization—rooted in emotional and social hostility—has become particularly salient. Political opponents are increasingly seen not as legitimate adversaries but as illegitimate enemies, contributing to the degradation of democratic discourse. This dynamic is further amplified by populist rhetoric, which reinforces in-group identity while exacerbating out-group animosity, creating a political climate in which compromise is framed as betrayal \cite{Mudde2017, Iyengar2009}.

Along the same line, the European Union (EU) combines both features at the European Parliament level, with the proliferation of Eurosceptic and populist parties \cite{RipollServent2019} and the widening ideological rifts within traditional political groups \cite{Brzel2023} – contributing to the forces straining coalition-building and consensus-driven policymaking efforts within the EU and challenging legislative cohesion. The European Parliament’s structure, characterized by a balance of ideological alignment and national interests, offers a valuable case study for examining polarization within a multinational legislative framework \cite{Hix2009, Lo2018}. 

Within the context of the European Parliament, these cleavages manifest not only between political groups but also within them, as MEPs balance their ideological affiliation with national imperatives. This dual allegiance adds complexity to coalition formation and challenges the assumption of stable, ideologically cohesive voting blocs. Recent studies have shown that while a dominant left-right axis structures much of the EP's voting behavior, a second, cross-cutting dimension—centered on attitudes toward European integration—frequently disrupts expected patterns of alignment, especially as Eurosceptic factions gain ground \cite{Brzel2023, Lo2018}.

While substantial research has explored polarization in national parliaments, such as in the United States \cite{Canen2020}  and the United Kingdom \cite{Peterson2018, Evans2017}, a gap remains in our understanding of how these phenomena unfold in a transnational legislative body like the European Parliament \cite{Hix2009, Brzel2023}. Empirical evidence on the evolution of polarization in the European Parliament, particularly concerning voting behavior across distinct legislative subjects, remains limited, underscoring the need for a more in-depth investigation into these dynamics. In particular, the patterns of synergy and antagonism formed between the participant actors and how they contrast with different subjects.

Here, we utilize roll-call data on Members of the European Parliament (MEPs) to analyze the co-voting network for the 9th term of the European Parliament (2019-2024). This type of data (roll-call) records how each MEP voted (whether in favor, against, or abstained) on individual legislative proposals. Moreover, we examine the alliances and divisions among political groups on major legislative subjects. By mapping patterns of co-voting behavior, we demonstrate that ideological alliances in the European Parliament are highly contingent on legislative subject matters, often transcending group boundaries and giving rise to dynamic, multi-polar voting communities that challenge conventional notions of stable group alliances.

\section{Data}
We use roll-call data on MEPs’ votes in the 9th term (2019-2024) of the European Parliament, sourced from the HowTheyVote.eu \cite{HowTheyVoteeu2022} platform and supplemented with legislative subject information from the Legislative Observatory of the European Parliament \cite{EuropeanParliament1994}. 

A bill in the European Parliament refers to a legislative proposal, motion, or report that is put up for debate and a formal roll-call vote. Each bill is assigned to one or more legislative subjects, following a three-level classification system established by the European Parliament's Legislative Observatory. The top level, referred to as primary subjects, consists of eight broad policy areas:
\begin{itemize}
    \item European citizenship – Covers matters of citizenship, migration, and asylum.
    \item Internal market, single market – Internal market regulations, competition, and consumer rights. It also covers standards for products, public subsidies, and the mutual recognition of degrees and qualifications.
    \item Community policies – European political parties and the European Parliament’s interaction with other institutions, on the European and national levels. It also covers the implementation of European treaties and all general issues regarding the EU institutions.
    \item Economic, social and territorial cohesion – Employment, social, and infrastructural policies, all of which aim to promote the EU’s economic, social, and territorial cohesion.
    \item Economic and monetary system – The economic and monetary system of the EU and regulations regarding the financial system, taxation, competition, and the free movement of capital and payments. It also comprises the relationship with the European Central Bank (which is accountable to the European Parliament).
    \item External relations of the Union – Relationship of the EU with third countries. Decisions concerning the accession process to the EU, associations, and the European Neighborhood Policy fall into this category. Security, defense, development policy, and human rights matters are also part of this policy area.
    \item Area of freedom, security and justice – Home affairs and policies regarding justice and the freedom of movement within the EU. It also covers harmonizing legal systems and cooperation at the police and justice levels between member states. The Schengen Area, the European Arrest Warrant, and Frontex patrols are well-known initiatives associated with this policy area.
    \item State and evolution of the Union – European integration process and concerns the power balance between the EU institutions on the one side and the member states on the other.
\end{itemize}

\begin{figure}[t]
    \centering
    \includegraphics[width=\textwidth]{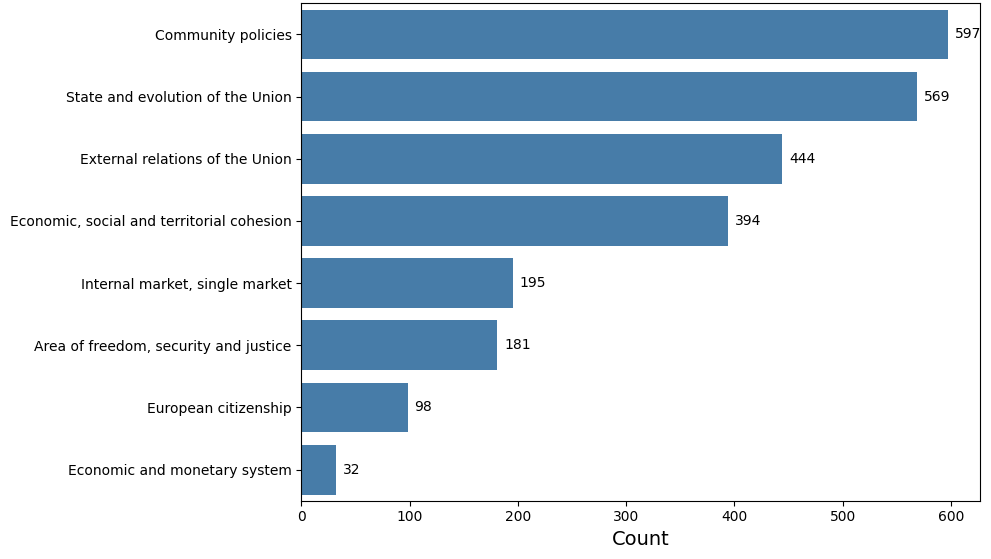}
    \caption{Distribution of bills by primary subjects during the 9th term of the European Parliament}\label{fig0}
\end{figure}

The bills are voted on by nationally elected MEPs from the 28 EU member states (including the United Kingdom until January 31, 2020). MEPs have three possible voting positions: “FOR”, “AGAINST”, or “ABSTENTION”. However, if an MEP was not serving during the vote or was not present in the voting session, their position is recorded as “DID\_NOT\_VOTE”.

MEPs join one of several transnational parliamentary groups based on ideological alignment rather than nationality. During the 9th term of the European Parliament, there were seven recognized political groups: EPP, S\&D, Renew Europe (Renew), The Greens/European Free Alliance (Greens/EFA), European Conservatives and Reformists Group (ECR), Identity and Democracy (ID), and The Left in the European Parliament (The Left). In addition, some MEPs remained non-attached (NI), meaning they did not affiliate with any political group.

Approximately 32.6\% of the MEPs did not serve the entire term, indicating a substantial portion with limited legislative activity. Hence, to ensure our study is based on MEPs with significant voting activity and reduce distortions from MEPs who served for too little time, we follow the approach of Schoch \& Brandes \cite{Schoch2020} and exclude MEPs who participated in fewer than 50\% of all roll-call votes. 

Voting positions were binarized into two classes: “supporting” that includes the “FOR” votes, and “not\_supporting” that groups the “AGAINST”, “ABSTENTION”, and “DID\_NOT\_VOTE” votes. Grouping the remaining three positions into “not\_supporting” partially mitigates the data imbalance (see Figure 2) but, more importantly, ensures that the data is suitable for the co-voting analysis in the next steps.

\begin{figure}[t]
\centering
\includegraphics[width=\textwidth]{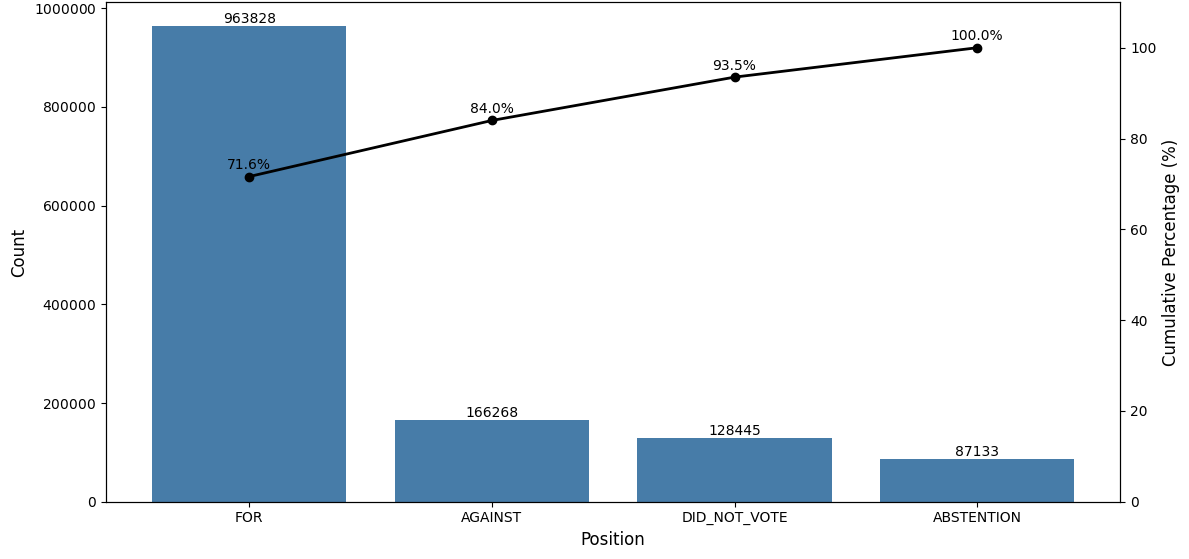}
\caption{Distribution of voting positions during the 9th term of the European Parliament}\label{fig1}
\end{figure}

Next, following the steps from NOMINATE \cite{Poole1985}, we removed high-consensus bills. To that end, we excluded any bill on which fewer than 2.5\% of MEPs opposed the majority position. For example, if 98\% of MEPs voted in support of a bill and only 2\% did not, we would consider the bill too consensual. This step ensures that our study focuses only on bills with minimal group divisiveness, reducing the noise introduced by overly consensual bills.

Furthermore, some MEPs appear in multiple groups or as NI at different times. Although NI is not a recognized group, we treat it as one for the sake of a single membership assignment. This means that each MEP will be assigned to only one group (or NI) throughout the study, even if they held multiple affiliations during the term. To determine this single “most representative” affiliation, we examine all group (or NI) memberships every MEP held and the time spent in each. Whichever affiliation occupied most of the MEP’s term is chosen. If there is a draw in terms of time, we select the first affiliation that the MEP held. This approach avoids representing the same MEP multiple times with different groups (or as NI). Consequently, each MEP is allocated exactly one affiliation, whether a recognized political group or NI, ensuring a streamlined yet accurate depiction of each MEP’s predominant membership.

Finally, we excluded all MEPs who spent more time as NI than as affiliated with any group. Since those MEPs lack significant group affiliation, their voting behavior does not contribute meaningfully to our study. The final working dataset provides information on the voting patterns of 695 MEPs and seven political groups on 1,890 bills, representing a total of 1,273,626 roll-call votes.

\section{Co-Voting Networks}
The next step involves inferring the co-voting networks. In such networks, nodes represent MEPs and edges connect pairs of MEPs who exhibit statistically significant co-voting behavior. To that end, following the steps outlined in references \cite{Domagalski2021, Saracco2015}, we employed the \textit{Bipartite Configuration Model} (BiCM) to extract the backbone network from the co-voting data.

We begin by encoding the roll-call data as a biadjacency matrix $B \in \{0,1\}^{m \times b}$, where $m$ represents the number of MEPs and $b$ represents the number of bills. Each element $B_{ik}$ equals one if MEP $i$ supported bill $k$ and is zero otherwise. Such a matrix effectively represents a bipartite network that links MEPs to bills they have supported. Then, we compute the weighted unipartite projection $P = BB^T$, where each entry $P_{ij}$ reflects the number of bills co-supported by MEPs $i$ and $j$, while the diagonal $P_{ii}$ gives the total number of bills supported by MEP $i$.

To extract only the statistically meaningful co-supporting ties, we need to compare the observed $P_{ij}$ with the expectations under a null model that preserves the observed support frequencies of each MEP (row sums of $B$) and the popularity of each bill (column sums of $B$). To that end, the BiCM formalizes this null model using the principle of maximum entropy: it generates an ensemble $\varepsilon = \{ B^* \}$ of random bipartite matrices, where the probability of an edge between MEP $i$ and bill $k$ is:
\begin{equation}
p_{ik} = \frac{x_i y_k}{1 + x_i y_k}
\end{equation}
where $x_i$ and $y_k$ are positive real-valued parameters associated with MEP $i$ and bill $k$, respectively. These parameters are not directly observed but fitted numerically to ensure that the expected degrees in the model match the observed ones:
\begin{equation}
  \sum_k p_{ik} = R_i \quad \text{and} \quad \sum_i p_{ik} = C_k  
\end{equation}
where $R_i$ is the total number of bills supported by MEP $i$ and $C_k$ is the number of MEPs who supported bill $k$. Thus, $x_i$ and $y_k$ are latent factors that encode each MEP’s support tendency and the overall popularity of each bill. 

\begin{figure}[!t]
\centering
\includegraphics[width=\textwidth]{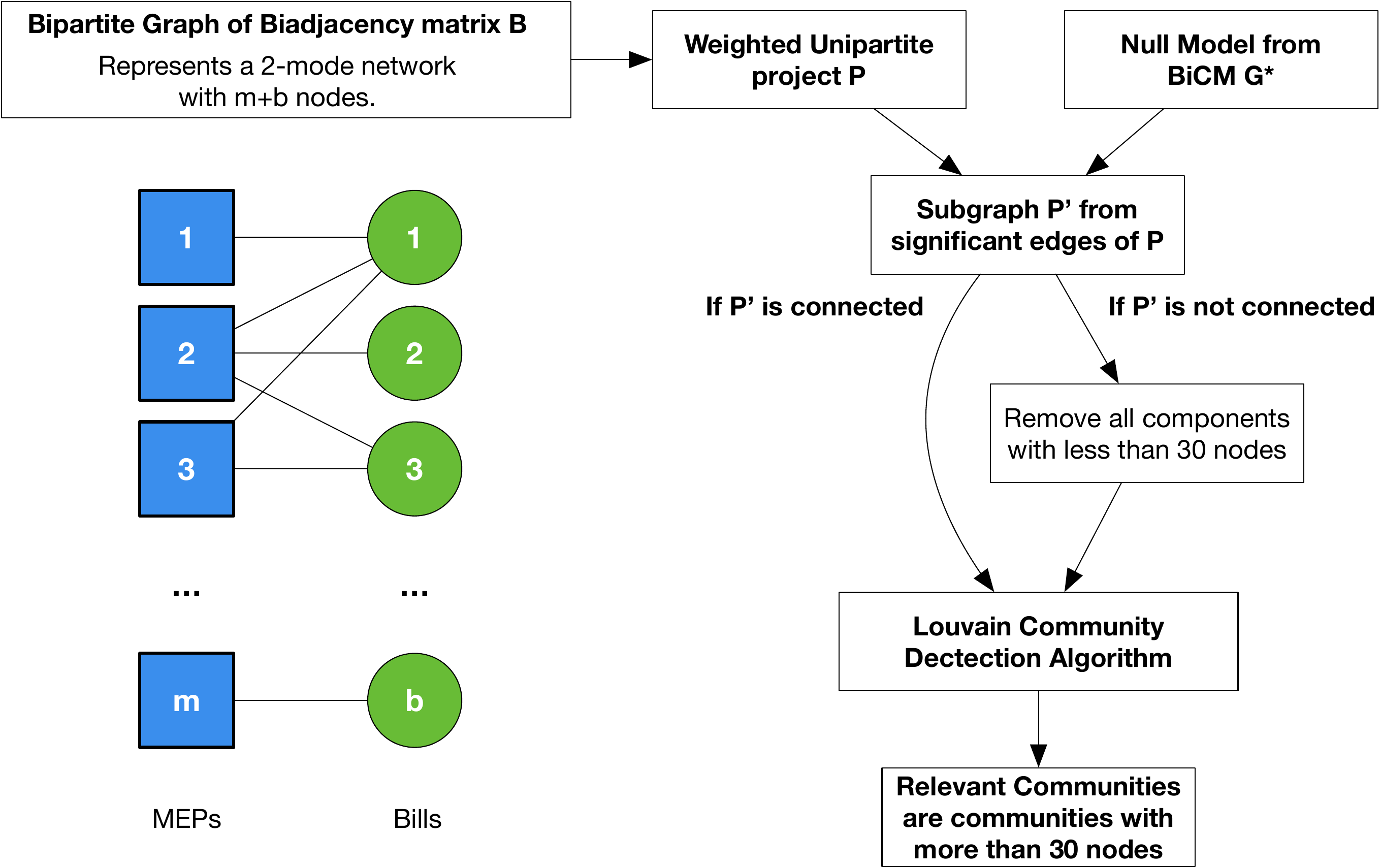}
\caption{Methodology workflow with the steps from the source data until obtaining the relevant communities.}\label{fig1}
\end{figure}

Hence, each synthetic matrix $B^* \in \varepsilon$ is a realization of the null model, where the entries $B^*_{ik} \sim \text{Bernoulli}(p_{ik})$ are independent. We then project $B^*$ onto the MEP space via $G^* = B^* B^{*T}$, analogous to the observed projection $P$. For each MEP pair $(i,j)$, the entry $G^*_{ij} = \sum_{k=1}^b B^*_{ik} B^*_{jk}$ counts the number of bills that $i$ and $j$ support jointly in a single draw from the null model. Because $B^*_{ik} \sim \text{Bernoulli}(p_{ik})$ and the pairs $(B^*_{ik}, B^*_{jk})$ are independent across $k$, the product $B^*_{ik} B^*_{jk} \sim \text{Bernoulli}(p_{ik} p_{jk})$. Therefore, $G^*_{ij}$ follows a Poisson Binomial distribution with parameters $\{ p_{ik} p_{jk} \}$, which models the sum of $b$ independent (but not identically distributed) Bernoulli trials.

We then perform a right-tailed hypothesis test for each MEP pair $(i, j)$ to determine whether their observed co-support count $P_{ij}$ is significantly higher than expected under the null model. Formally, we compute the p-value:
\begin{equation}
\pi_{ij} = P_{\varepsilon}(G^*_{ij} \geq P_{ij})
\end{equation}
which measures the probability that two MEPs would co-support at least $P_{ij}$ bills purely by chance. The null hypothesis posits that the co-support between MEPs $i$ and $j$ arises entirely from random alignment, given their marginal support levels and the overall popularity of each bill. That is, $i$ and $j$ appear to vote similarly only because they each support many bills and not because of any ideological proximity. Rejecting the null hypothesis implies that the observed co-support between MEPs $i$ and $j$ is too frequent to be explained solely by these marginal effects.

We retain an edge in the backbone network $P'$ if $\pi_{ij} < \alpha$. We set the significance level $\alpha$ to 0.01 to minimize false positives without compromising the backbone's sparsity. This results in an unweighted backbone network $P' \subset P$, which includes only co-support relationships that are statistically stronger than expected under the BiCM null model. All computations concerning the BiCM were performed using the R package \texttt{backbone} \cite{Neal2022}.

\section{Results \& Discussion}
We begin by applying the co-voting network construction method to aggregate all subjects, as well as to each primary subject independently, to identify the relevant voting communities. During this process, we observed that the primary subject “Economic and Monetary System” contained only 31 bills, and therefore, it was not studied. Figure~\ref{fig1} illustrates the steps for obtaining the relevant subgraphs and communities.

\begin{figure}[!t]
\centering
\includegraphics[width=\textwidth]{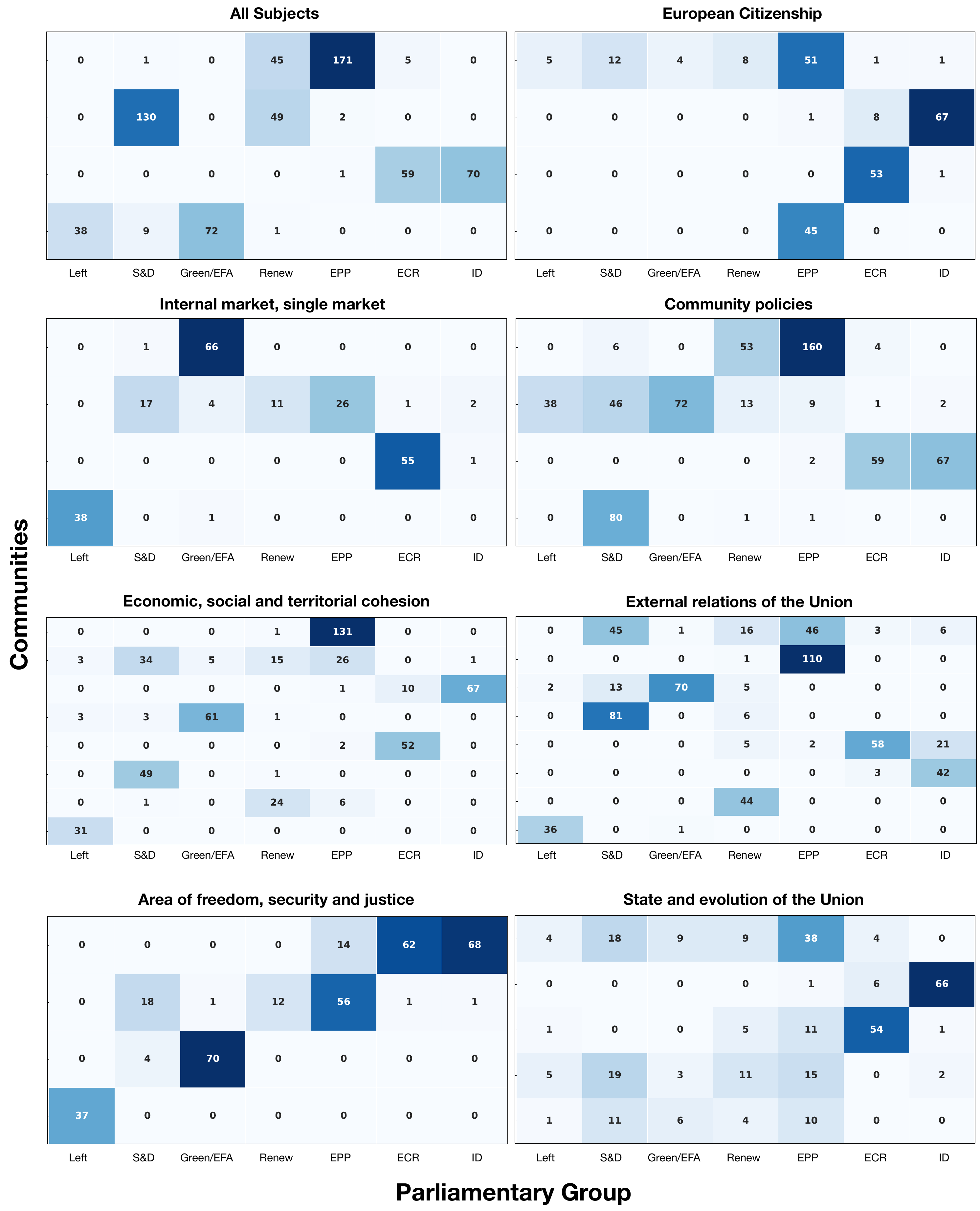}
\caption{Profiling of obtained communities according to their composition regarding Parliamentary Groups.}\label{fig2}
\end{figure}

\begin{figure}[!t]
\centering
\includegraphics[width=0.9\textwidth]{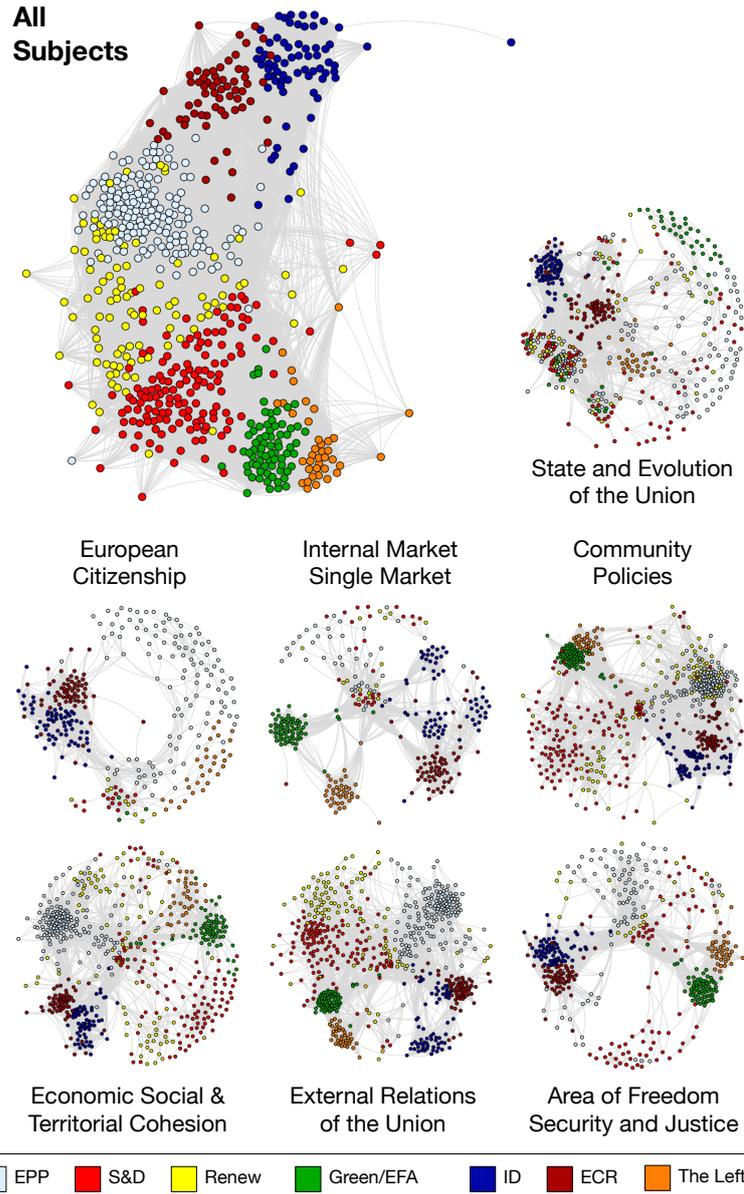}
\caption{Co-voting networks for all subjects (top left) and then for each of the primary subjects studied. Each node represents an MEP, colored according to the most representative Group affiliation, and links connect MEPs with a significant co-voting relationship. Network visualizations were generated using Wolfram Mathematica V13 with the GraphPlot function and using the "Gravity Embedding" Layout.
}\label{fig3}
\end{figure}

To each subgraph $P'$, we assess the connectivity. If connected, we applied the Louvain community detection algorithm \cite{blondel2008fast} and retained only communities with at least 30 notes (smaller communities were considered too noisy for meaningful analysis). If the subgraph was disconnected, we first removed all components with fewer than 30 nodes and then applied the Louvain community detection algorithm to the remaining network, again retaining only communities with at least 30 nodes, therefore obtaining the relevant voting communities  The outcome of the relevant voting community detection process, applied both to the aggregation of all subjects and to each primary subject individually is summarized in both Figure~\ref{fig2} and Figure~\ref{fig3}.

Consistently across most subjects, Greens/EFA and The Left tend to be highly concentrated within single communities, suggesting a high degree of ideological cohesion and coordinated voting behavior. In contrast, centrist and governing groups -- such as Renew, EPP, and S\&D -- often exhibit fragmented patterns, with their MEPs distributed across multiple voting communities. This fragmentation suggests that their positions vary substantially across different policy domains, confirming the issue-driven nature of alliances and the erosion of traditional centrist bloc voting.

In the case of “State and evolution of the Union,” a clear fragmentation emerges along the pro- and anti-European integration axes. The Eurosceptic groups, ID and ECR, form their own distinct and internally cohesive communities, indicating strong alignment in opposing deeper EU integration. However, their separation into different communities suggests that they embody divergent strands of Euroscepticism rather than a unified bloc. Similarly, the Euroenthusiastic groups – EPP, Renew, and S\&D – are distributed across three different communities, revealing underlying factionalism within the pro-integration camp.

Overall, the co-voting networks substantiate the broader analytical claim of this research: the European Parliament exhibits multi-polar co-voting behavior driven by a combination of ideological affinity, subject specificity, and the emergence of a Eurosceptic-Euroenthusiast cleavage that challenges stable coalition dynamics.

For all subjects and each primary subject independently, we complement the analysis on the subgraph $P'$ (seen in Figure~\ref{fig1}) with several network-based statistics: optimal modularity; group modularity; nationality modularity; group assortativity; nationality assortativity; average clustering coefficient; network density; percentage of remaining nodes; total detected communities; and total relevant communities.

\begin{table}[]
\footnotesize
\centering
\caption{Modularity and Assortativity groups across different subjects. Optimal modularity was identified using the Louvain algorithm, which represents an upper structural limit for each graph and thus serves as the reference for "optimal" modularity. We compare such modularity to the Modularity when assuming graph partitions along the MEPs' Parliament groups' affiliations and nationalities. Moreover, assortativity measures the likelihood that MEPs tend to form voting ties with peers of the same attribute (here, the attribute can be group or nationality), rather than forming cross-attribute alliances. We measured assortativity in terms of the MEP European Parliamentary group and the nationality of its members.}
\begin{tabular}{lccccc}
\hline
\rowcolor[HTML]{EFEFEF} 
\cellcolor[HTML]{EFEFEF} &
  \multicolumn{3}{c}{\cellcolor[HTML]{EFEFEF}\textbf{Modularity}} &
  \multicolumn{2}{c}{\cellcolor[HTML]{EFEFEF}\textbf{Assortativity}} \\
\rowcolor[HTML]{EFEFEF} 
\multirow{-2}{*}{\cellcolor[HTML]{EFEFEF}\textbf{Subject}} &
  \textbf{Optimal} &
  \textbf{Group} &
  \textbf{Nationality} &
  \textbf{Group} &
  \textbf{Nationality} \\ \hline
All subjects                              & 0.520 & 0.406 & 0.041                         & 0.512 & 0.044 \\
European citizenship                      & 0.483 & 0.412 & 0.179 & 0.625 & 0.211 \\
Internal market, single market            & 0.718 & 0.609 & 0.188                         & 0.787 & 0.210 \\
Community policies                        & 0.553 & 0.413 & 0.067                         & 0.553 & 0.073 \\
Economic, social and territorial cohesion & 0.662 & 0.549 & 0.124                         & 0.740 & 0.137 \\
Economic and monetary system              & 0.803 & 0.484 & 0.549                         & 0.991 & 0.717 \\
External relations of the Union           & 0.715 & 0.605 & 0.127                         & 0.736 & 0.138 \\
Area of freedom, security and justice     & 0.607 & 0.522 & 0.108                         & 0.663 & 0.121 \\
State and evolution of the Union          & 0.586 & 0.350 & 0.080                         & 0.423 & 0.087 \\ 
\hline
\end{tabular}
\end{table}

The analysis of co-voting behavior in the 9th term of the European Parliament reveals a legislative landscape marked not by binary ideological bifurcation, but by multi-polarization. This reality is evident across the legislative agenda, and it is particularly pronounced in specific legislative subjects. These findings challenge prevailing assumptions derived from two-party systems, such as those of the United States or the United Kingdom, where polarization is predominantly conceptualized along a single left-right axis.

When considering all subjects together, the optimal modularity of 0.520 and group modularity of 0.406 suggest that MEPs generally form cohesive voting blocs along ideological lines, though not rigidly. The small gap between optimal and group modularity indicates that group affiliation is the main, but not exclusive, driver of voting behavior. Cross-group coalitions and issue-specific preferences introduce meaningful variation, as reflected in the moderate group assortativity of 0.512. In contrast, national affiliations play a marginal role, as evidenced by the low modularity (0.041) and assortativity (0.044) of nationality. This confirms that MEPs rarely vote in nationally cohesive blocs, underscoring the European Parliament’s transnational nature.

For the all subjects networks, statistics confirm a cohesive yet multi-polar European Parliament. A high clustering coefficient (0.654) indicates strong within-group ties, while a low density (0.190) suggests sparse links across communities. Four sizeable communities emerge, not just two, and the backbone still contains approximately 95\% of MEPs across 1890 bills, evidence that these patterns are both clear and robust. 

The voting communities reflect identifiable patterns for all subjects. The EPP and S\&D remain consistently separated, confirming an ideological divide between center-right and center-left forces. Renew demonstrates a bifurcated alignment pattern, affiliating at times with S\&D and at other times with EPP, mirroring its centrist position and ideological flexibility. The Eurosceptic and far-right groups, ECR and ID, form an isolated voting bloc. On the opposite pole, Greens/EFA and The Left coalesce.

\begin{table}[]
\centering
\caption{\textbf{Structural Statistics of the co-voting networks}. The average clustering coefficient (CC) offers a means to quantify the degree to which MEPs form tightly interconnected voting clusters. The network density (Density) measures the proportion of connections in relation to the theoretical maximum given the number of MEPs. The Remaining Nodes presents the percentage of the population of MEPs that are represented in each network. $N_c$ represents the number of identified communities. $\Tilde{N}_c$ is the number of communities with at least 30 MEPs, which we refer to as the relevant communities.
}
\footnotesize
\begin{tabular}{lcccccc}
\hline
\rowcolor[HTML]{EFEFEF} 
\textbf{Subject} &
  \textbf{\begin{tabular}[c]{@{}c@{}}CC\end{tabular}} &
  \textbf{\begin{tabular}[c]{@{}c@{}}Density\end{tabular}} &
  \textbf{\begin{tabular}[c]{@{}c@{}}\% Remaining \\ Nodes\end{tabular}} &
  \textbf{\begin{tabular}[c]{@{}c@{}}$N_c$\end{tabular}} &
  \textbf{\begin{tabular}[c]{@{}c@{}}$\Tilde{N}_c$\end{tabular}} &
  \textbf{\begin{tabular}[c]{@{}c@{}}Number\\ of bills\end{tabular}} \\ \hline
All subjects                              & 0.654 & 0.190 & 94.964\% & 5  & 4 & 1890 \\
European citizenship                      & 0.485 & 0.062 & 46.619\% & 14 & 4 & 98   \\
Internal market, single market            & 0.659 & 0.066 & 57.410\% & 39 & 4 & 191  \\
Community policies                        & 0.671 & 0.117 & 93.957\% & 8  & 4 & 595  \\
Economic, social and territorial cohesion & 0.607 & 0.060 & 88.345\% & 25 & 8 & 386  \\
Economic and monetary system              & 0.715 & 0.059 & 12.374\% & 20 & 0 & 31   \\
External relations of the Union           & 0.620 & 0.063 & 93.957\% & 11 & 8 & 444  \\
Area of freedom, security and justice     & 0.577 & 0.065 & 73.237\% & 49 & 4 & 176  \\
State and evolution of the Union          & 0.621 & 0.073 & 77.554\% & 40 & 5 & 567  \\ \hline
\end{tabular}
\end{table}

These dynamics are accentuated when legislative activity is disaggregated by primary subject. In nearly every primary subject, the optimal modularity surpasses the value for all subjects, indicating that multi-polarization tends to deepen when policy specificity increases. For instance, the terms “Internal market, single market” and “External relations of the Union” exhibit optimal modularities of 0.718 and 0.715, respectively, indicating the emergence of sharply defined voting blocs within these subjects. This trend is mirrored by increases in group modularity and assortativity across most primary subjects, suggesting that group affiliations become more determinative when MEPs are engaged in focused policy deliberation.

An exception is the “State and Evolution of the Union” primary subject, where group modularity and assortativity drop significantly. This suggests that traditional ideological cleavages lose explanatory power, giving way to a new axis of conflict: Eurosceptics versus Euroenthusiasts. Voting communities reflect this shift, with ECR, ID, and a few EPP dissenters forming distinct Eurosceptic blocs, while Euroenthusiastic groups (EPP, S\&D, and Renew) cluster separately. However, the fact that these pro-integration groups form distinct communities reveals underlying factionalism. These patterns reinforce the view that attitudes toward European integration now constitute a second dimension of conflict in the European Parliament.

Though limited overall, the role of nationality gains more salience within primary subjects. In every primary subject, nationality modularity and assortativity exceed the baseline observed across all subjects. This suggests that while nationality is not a primary driver of voting behavior, it becomes more relevant in context-specific legislative matters, particularly when domestic political considerations become more acute.

These findings reinforce the idea that the European Parliament’s polarization is not merely a function of increasing ideological distance between stable partisan camps but a reflection of complex, policy-contingent multipolarity. Legislative consensus in the European Parliament thus depends not on fixed grand coalitions but on navigating a fluid, multi-polar alliance space.

\section{Conclusions}
This research aimed to examine the structure of political multipolarization within the European Parliament during its 9th term (2019-2024), characterized by profound social, economic, and geopolitical disruptions. Drawing on co-voting networks derived from roll-call data, this research examined how MEPs aligned or diverged across subjects. The findings challenge narratives of polarization in supranational institutions and contribute several novel insights to the political science and network analysis literature.

The most salient contribution of this research is the empirical evidence of multi-polarization within the European Parliament. Contrary to the dominant literature \cite{McCarty2016, Canen2020, Peterson2018}, rooted mainly in the binary political systems of the United States and United Kingdom, this research demonstrates that the ideological structure of the European Parliament is not reducible to a single left-right dimension. In subjects such as “Community policies”, “Internal market, single market”, and “External relations of the Union”, ideological alliances shift significantly, reflecting fluid coalitions rather than static group blocs. We demonstrate that relationships in these communities appear to align preferentially by ideological affinity and group membership, rather than by nationality. These findings reaffirm the European Parliament's view as a transnational legislature in practice, not merely in structure. In this context, a particularly striking result is that some of the most cohesive voting behavior is observed within the Greens/EFA and The Left, while traditional governing groups – EPP, S\&D, and Renew – exhibit lower internal cohesion and participate in multiple, sometimes conflicting, voting alliances. This suggests an erosion of uniform group discipline within centrist groups, highlighting a shift toward more flexible, issue-driven alignments. 

Moreover, the subject “State and evolution of the Union” reveals a novel axis of multi-polarization: Euroscepticism versus Euroenthusiasm. This axis cuts across traditional left-right divides. This cleavage represents a redefinition of political conflict within the European Parliament and may presage the emergence of a second ideological dimension that will dominate future EU policymaking.

Taken together, these findings redefine our understanding of polarization in supranational governance. The European Parliament, far from being a static or derivative arena of national political conflict, emerges instead as a dynamic and uniquely structured legislative body, where polarization manifests not as a fixed binary but as a multidimensional and fluid network of issue-driven alliances and cleavages. However, several limitations might constrain these conclusions. First, the number of bills per subject varies widely, which may reduce the reliability of results for subjects with fewer bills. Second, by focusing on only one legislative term, our dataset yields sparser communities for subjects that are less represented; therefore, the interpretation of their communities must be treated with caution. Third, for detecting relevant communities and computing optimal modularity, we relied solely on the Louvain community detection algorithm. A more robust approach would combine multiple community detection algorithms and aggregate the results.

\backmatter


\bmhead{Data Statement}
The datasets used and/or analysed during the current study are available from the corresponding author on reasonable request.


\bmhead{Authors Contributions}
SMR (Conceptualization, Methodology, Data Curation, Formal Analysis, Writing - Original Draft), AC ( Writing - Review \& Editing, Supervision), FLP (Writing - Review \& Editing, Supervision).


\bmhead{Acknowledgements}
This work was supported by national funds through FCT (Fundação para a Ciência e a Tecnologia), under the project UIDB/04152 - Centro de Investigação em Gestão de Informação (MagIC)/NOVA IMS and under the project 2024.07378.IACDC (https://doi.org/10.54499/2024.07378.IACDC). The authors are grateful to the HowTheyVote.eu team for sharing the data and their prompt interest in addressing our questions.


\bibliography{references.bib}
\end{document}